\documentclass[12pt,nofootinbib]{revtex4-1}
\usepackage{epsfig}
\newcommand{\ave}[1]{\left\langle #1 \right\rangle}

  \newcommand{\lqcd}{\Lambda_{\mathrm{QCD}}}
 
\newcommand{\eqcomma}{\phantom{AA},\phantom{AA}}

\newcommand{\order}[1]{ \mathcal{O} \left( #1 \right) }

\usepackage{graphicx}

\begin{document}
\title{Multi-particle correlations, many particle 
systems, and entropy in effective field theories}
\author{Giorgio Torrieri}
\affiliation{FIAS,
  J.W. Goethe Universit\"at, Frankfurt A.M., Germany }
\affiliation{Pupin Physics Laboratory, Columbia University, 538 West 120$^{th}$ Street, New York,
NY 10027, USA\\
torrieri@phys.columbia.edu}
\begin{abstract}
We discuss the treatment, in an effective field theory,  of multi-particle correlations within a ``large'' system.   We show that the act of coarse-graining necessarily introduces violations of unitarity in the evolution of states where the particle number is not defined.
For an interacting system, such unitarity violations can cascade from the ultraviolet scale to the infrared in a ``short'' time.
Hence, an effective field theory will be grossly inadequate for describing multi-particle correlations and related observables, even far away from the fundamental scale $\Lambda$.   We furthermore argue that if the system is strongly coupled at $\Lambda$, than its final state {\em in the Effective Field Theory} (EFT) will appear as the highest entropy state if only low cumulants and correlations of the EFT degrees of freedom are measured.
   Heuristically, this can serve as an explanation of how ``entropy'' is created in a microscopically unitary evolution of a Quantum Field Theory (QFT).
We conclude by discussing how these considerations might provide a clue to the apparent thermalization  in a hadronic collision even in comparatively small systems, as well as the so-called black hole information paradox;   We argue the ``paradoxes'' are likely to be artifacts of using an effective theory beyond its domain of validity.
 \end{abstract} 
\maketitle 
\section{Introduction \label{secintro}}
Quantum field theory, while being extremely successful from a phenomenological point of view, still has profound fundamental issues with its mathematical definition.
It is well known \cite{haag1,haag2} that the interaction picture is inexorably ill-defined for an interacting quantum field theory, since no unitarily equivalent mapping between a ``free theory'' and an ``interacting theory'' can exist.   Hence, what is called a ``free particle'' in an interacting theory differs from a real ``free particle'' to an extent that any matrix elements between these two theories generally vary by infinite amounts, even in the perturbative limit.

An effective way to deal with this problem is renormalization.   ``Fast'' degrees of freedom, with momenta $k_i$ bigger than some ``fundamental'' scale $\Lambda$ much larger than the typical momentum exchange between particles in a given experiment, are integrated over.  For renormalizeable terms, the shifts, of order $\sim \Lambda$, can be absorbed into parameters of a Lagrangian written solely in terms of ``slow'' terms, so $\Lambda$ becomes physically irrelevant until such high energies that $k \sim \Lambda$ are reached.
For non-renormalizeable theories, this cannot be done.    However, an effective theory power series can be defined in which $k/\Lambda$ is a small parameter.
Such an ``effective Lagrangian'' will hopefully have only a few leading terms incorporating all the relevant quantum fluctuations.
While renormalization was initially considered to be a dubious ``sweeping infinities under the carpet'', it is now understood to be a conceptually elegant implementation of the fact that nature seems to separate energy scales (you do not need quantum gravity to calculate the trajectory of an apple falling from a tree!) and therefore a consistent theory should exist where only ``relevant'' degrees of freedom are considered.  While we have to live with the fact that observables at low energy cannot uniquely specify the theory at high energy, the fundamental problems of QFT become ``harmless'' for low energies.  We can therefore hope they will be fixed a UV quantum theory of gravity\cite{intro} which lives at $\Lambda =E_p=10^{19}$GeV, safely out of reach of our current experiments.

In elementary particle experiments, this approach is particularly effective because our observables are definite $n-$particle scattering matrices, $<k_1^i k_2^i ... k_i^i | k_1^f k_2^f ... k_j^f>$.    The detector can distinguish events such as $e^+ e^- \rightarrow \mu^+ \mu^- \gamma$ from, say, $e^+ e^- \rightarrow \mu^+ \mu^- \gamma \gamma e^+ e^-$, and measure relative probabilities of such events occurring given a controlled flux of $e^+ e^-$ at definite $\sqrt{s}$.
Events with ``too many particles'' are forbidden by kinematics.
This is not the situation for a ``macroscopic object'', with ``many'' particles, even if the typical energy of each particle is comparatively low.   A plasma (whether in the early universe, or in heavy ion collisions), a strong ``classical field'', and most likely a macroscopic black hole are examples of such systems.

Here, the power expansion described earlier breaks down:  While each momentum $k_i \ll \Lambda$, it is not true that $\ave{N k_i}\ll \Lambda$.  
Some care therefore is needed in applying the concepts of renormalization theory to many-particle systems.   

This theoretical uncertainity gives rise to some ``phenomenological puzzles'' under active investigation:  
Perhaps the most well-known is the ``black hole information problem'' \cite{bhinfo}, the apparent violation of unitarity in the process of converting a ``quantum pure state'' into Hawking radiation via the creation of a black hole.
What makes this seem like a paradox is that, while we do not know much about the regime valid at $E_p$, the effective field theories ``long before'' a black hole's formation and ``long after'' its evaporation, as well as on the horizon, are very well known (they are to a good approximation, free field theories in flat space).
While of course this problem cannot be experimentally investigated, analogues of it might be accessible to experiment \cite{adscft1,adscft2,adscft3,adscft4,nastase}:
 Hadronic collisions give rise to a ``fireball'' whose dynamics are regulated by strongly coupled QCD, whose dynamics are not understood at the quantitative level.   We think, however, that the final state at $T\ll \lqcd$, is very well understood as a quasi-free gas of pions and protons.   Indeed, entropy and other thermal concepts are readily linked to observational counting of such free degrees of freedom   ~\cite{Fer50,Pom51,Lan53,Hag65,pbm,jansbook,becattini}.   In fact, the increase of entropy at deconfinement was thought to be a QGP signature \cite{entroexp}.
But this is also a unitary quantum process, whose UV theory (QCD) is known to be an ``ordinary'' quantum field theory.   Where is the negentropy\cite{gibbs}\footnote{The ``negative entropy'' counted from the maximum entropy state.   It coincides with the ``capacity for entropy increase'' defined in \cite{gibbs}. At zero negentropy, a system is in thermal equilibrium,at maximum negentropy it is in a pure state} canceling out the particle entropy hiding?  And does this provide a key to the more abstract black hole puzzle, as, for example, Gauge/gravity duality \cite{adscft1,adscft2,adscft3,adscft4} makes us believe?   While several proposals were advanced recently to resolve these issues, no consensus exists as to in what limits does a unitary quantum process exhibit thermal-like behavior (see for example \cite{becattini,thermal1,thermal2}, or the difference in interpretation between \cite{becattini} and \cite{pbm,pbm2}).

 These notes are intended to clarify some of these issues, and discuss their implication to some problems of current research interest.   We examine the general problem of describing multi-particle systems, first the well-known classical analogue, and then in a quantum field theory.    We show where the scale separation problem might break down, and what effects it will have on observables, particularly the ``entropy'', observed and modeled in the infrared regime.   We finish by discussing the relevance of this discussion to concrete problems.

\section{Some intuition from classical physics: The BBGKY hierarchy \label{secclassical}}
Classically, the analogue of entropy conservation is Liouville's theorem \cite{classical1,classical2}:
A system of $N$ degrees of freedom can be represented as a point in $6N-$dimensional phase space.  If we start, at a certain point in time, with an ``ensemble of systems'' covering a phase space volume with a probability density function $f(x_i,k_i)$, this volume will be unchanged as the systems evolve in time\footnote{This definition is somewhat confusing, since it refers to an ensemble of nearly identical systems rather than a single system.  
The same can be said for the definition of entropy used later.
Given that the frequentist and Bayesian views of probability give the same quantitative results, however, we can continue to use such a ``frequentist'' view of entropy for a single system.  }.

\begin{figure}[h]
\begin{center}
\epsfig{width=18cm,figure=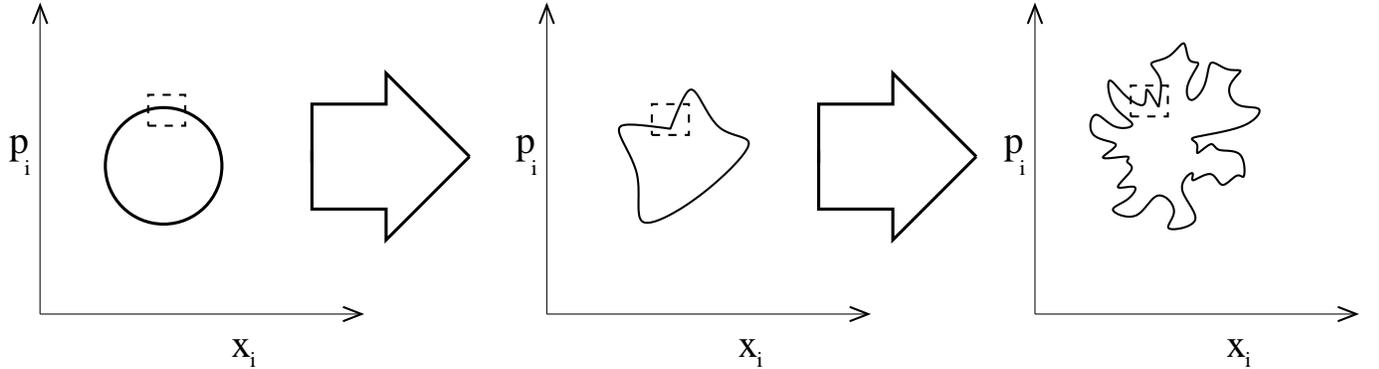}
\caption{\label{fractal} 1-particle phase space projection of a many-particle system as it evolves in time.   The dashed square denotes a limit of phase space, such as the one set by the uncertainity principle}
\end{center}
\end{figure}

Consider a system of $N$ classical particles\footnote{Liouville's theorem can break down when $N\rightarrow \infty$, something that arises in situations like ``shock heating'' in hydrodynamics, or first order phase transitions in the classical limit\cite{huang}.  These are most likely irrelevant to the current work, as both in this and the next sections the effective $N$ is large but finite }, interacting with a 2-particle translation invariant conservative potential $V_{ij}(|x_i-x_j|)$.
It can be shown \cite{bbgky1,bbgky2,bbgky3} that the classical s-particle Liouville equation for a system of $N$ particles is equivalent to the equation 
\begin{equation}
\label{bbgky}
  \sum_{i=1}^s \left( \dot{x_i} \frac{\partial f_s\left( x_1,x_2,...,x_N,k_1,k_2,...k_N \right)}{\partial x_i} - \sum_{j=1}^N \frac{\partial V_{ij}}{\partial x_i} \frac{\partial f_s \left( x_1,x_2,...,x_N,k_1,k_2,...k_N \right)}{\partial k_i} \right) =
\end{equation}
\[\ -\frac{\partial f_s\left( x_1,x_2,...,x_s,k_1,k_2,...k_s \right)}{\partial t} +
 \sum_{i=1}^s (N-s) \frac{\partial}{\partial k_i} \int \frac{\partial V_{i s+1}}{\partial q_i} f_{s+1}\left( x_1,x_2,...,x_{s+1},k_1,k_2,...k_{s+1} \right) dx_{s+1} dk_{s+1}
\]
this equation, easily generalizable to more than 2-particle potentials (provided translational symmetry and the conservative nature of the potential is preserved), encodes all the evolution of an $N-$particle distribution function defined on the ensemble of $6N$-dimensional phase space.   Its exact evolution will therefore respect Liouville's theorem on that space.    However, for more than $2-3$ degrees of freedom, and for pretty much any bounded Hamiltonian allowing for topological mixing in phase space ( i.e., a Hamiltonian free of issues like those discussed in \cite{ostro}), this equation's evolution will be highly chaotic:   Any ``neatly defined'' shape in phase space becomes more and more ``fractal'' with time (Fig. \ref{fractal}).
What this means is that more and more of the conserved phase space area becomes encoded in correlations between many particles:  If one particle had a ``very small'' deviation, after a logarithmically small time another particle will get a comparable deviation.     In this regime, equations such as Eq. \ref{bbgky} are inherently intractable\cite{classical1,classical2}.   They are also useless for all practical purposes, since in any case classical mechanics is thought to be at best an approximation of quantum mechanics, where phase space can only be known by a resolution set through the uncertainity principle, $\ave{(\Delta x_i)^2} \ave{(\Delta k_i)^2} \geq \hbar^2/4$.   When ``most of the area of phase space'' is encoded in details whose characteristic structure is smaller than this measure, the exact Liouville equation becomes useless to describe physical data.

Paradoxically, in this regime a {\em truncation} of the BBGKY series would do much better.
The virtue of equation \ref{bbgky} is that it can be thought of as an expansion in ever-decreasing $f_{s+1}$.    If cumulants and correlations above a critical $s$ are thought to be irrelevant (i.e., if we do not care about correlations between more fundamental degrees of freedom than $s+1$), Eq. \ref{bbgky} can be truncated into a system of $s$ non-linear differential equations.   If at $s \ll N$ one assumes
\begin{equation}
\label{truncs}
 \ave{ f_{s+1}\left( x_1,x_2,...,x_{s+1},k_1,k_2,...k_{s+1} \right)} \simeq c(\sigma) \ave{f_s \left( x_1,x_2,...,x_{s},k_1,k_2,...k_{s} \right)}\ave{ f_1(x_{s+1},k_{s+1})}
\end{equation} 
(or some more complicated factorization in terms of $f_i,f_j$ with $i+j=s$).
The ``noise coefficient'' $c(\sigma) \sim \sigma^s \ave{ \rho k }^{s/2}$ can be shown to be related,in a semi-classical picture, to the quantum mechanical scattering cross-sections \cite{kubo}.
For instance, in the Boltzmann equation $\ave{f(k_1,k_2,k_3,k_4)} \rightarrow \ave{f(k_1)}\ave{f(k_2)} \sigma_{k_1,k_2 \rightarrow k_3 k_4}$. 
 The Kadanoff-Baym equation \cite{kubo} can also be viewed as a higher-order extension of this approach.

One does not, however, expect such a system to obey Liouville's theorem and conserve phase space in its classical evolution.   Indeed, molecular chaos (truncating the BBGKY hierarchy at $s=1$) straight-forwardly leads to the Boltzmann H-theorem, showing entropy to be increasing \cite{bbgky1,bbgky2,bbgky3,bbgky4,bbgky5}.

While the exact derivation of such equations from quantum mechanics is still subject to some dispute, no one talks about the generation of entropy in truncated BBGKY systems as a paradox.  The increase of entropy is an artifact of the fact that any BBGKY truncation destroys inter-particle correlations.   The observability of such correlations, certainly in practice and even in principle, is dubious to impossible.   Hence, for ``all practical purposes'' Liouville's theorem ceases to apply and systems of many interacting particles generally thermalize.

Classical physics is now considered to be an ``effective theory'' of quantum theory, valid for when the ``total action'' $S_{total}\gg \hbar$.
Yet it is in exactly this limit, when the number of degrees of freedom $\gg 1$, that unitarity theorems that are exact both at the classical and quantum level (Liouville's theorem, quantum mechanically, is represented by unitarity) are violated.
A natural conclusion is that, in this limit, the effective expansion breaks down, even when the scale separation between the effective and fundamental theories is large.   In the rest of this work we shall argue this is indeed the case.  
\section{Effective theory and many-particle systems \label{seceft}}
\subsection{Effective field theories and renormalization: A short conceptual review \label{secrenoreview}}
The discussion in the previous section becomes a bit more subtle when a {\em relativistic quantum field theory} rather than a simple {\em many particle system} is considered.
In such a theory {\em any system} is a system of infinitely many off-shell particles, with the Fock space taking place of a Hilbert space.

Given a Lagrangian \cite{peskin}, for example a $\phi^4$ Lagrangian, ( the simplest Lagrangian of scalar particles, not carrying any conserved quantum numbers), one can define a ``bare'' action, 
\begin{equation}
\label{scalarfield}
S = -\frac{1}{2} \int d^3 x dt \left(  \partial_\mu \phi \partial^\mu \phi + m^2 \phi^2 + \frac{\lambda}{12}\phi^4 \right)
\end{equation}
When this action is inserted into the quantum partition function of the theory
\begin{equation}
Z(J) = \int \mathcal{D} \phi \exp \left[ i S(m,\lambda,\phi(x,t)) + J\phi(x,t) \right]
\end{equation}
and correlation functions $\ave{\phi(x_1)\phi(x_2)...\phi(x_n)} = \partial^n \ln Z(J)/\partial J^n$ are calculated beyond tree level, one finds that
for $\phi^4$ theory and for all interacting quantum field theories in four dimensions yields divergences appear.   The reason is that arbitrarily high momentum quantum fluctuations are allowed.   Ultimately, these divergences express the fact that arbitrary small differences between free and interacting theories yield infinite differences in observables \cite{haag1,haag2}.

The infrared behavior of this theory can however be regularized against some Ultraviolet cutoff scale, by performing truncations of some sort.   What this means in practice is that the quantum partition function of the theory
 is only sampled ``on a lattice'' of resolution below critical value $\Lambda^{-1}$.     If $\phi(x,t)$ is Fourier-expanded and separated into slow and fast componets via some function $R$ (approximating the $\Theta$-function)
\begin{equation}
\label{separation}
\phi(x,t) = \int d^3 k d E e^{i( k x-Et)}
 \left(  \phi_{slow}(k) R(k-\Lambda)R(E-\Lambda) +  \phi_{fast}(k)  \left[ \left(  1- R(k-\Lambda) \right)\left( 1-R(E-\Lambda)\right) \right] \right)
\end{equation}
 and only coefficients smaller than $\Lambda$ are kept:
\begin{equation}
\label{zj}
 Z(J) = \int \mathcal{D} \phi_{slow} \int \mathcal{D} \phi_{fast} \exp \left[ i S(\phi) + J\phi(x,t) \right]
\end{equation}
\[\ \underbrace{\simeq}_{x_i - x_j \gg \Lambda^{-1}} \order{\Lambda} \int \mathcal{D} \phi_{slow}  \exp \left[ i S'(\phi_{slow}) + J\phi_{slow} \right] \]
As long as we are interested in a finite number of correlation functions between fields, and the momenta of the interaction $\ll \Lambda$ the second equality holds, for any theory.    In addition, in renormalizeable interactions (such as the $\phi^4$ example) all residual dependence on $\Lambda$ can be hidden in a redefinition of parameters $m(\Lambda),\lambda(\Lambda)$ (in fact the exact form of the scale separator in Eq. \ref{separation} becomes irrelevant).     

The exact form of the equation describing the dependence of the Lagrangian parameters on the scale can be rigorously calculated in perturbation theory \cite{wilson} and non-perturbatively \cite{polch,wetterich} by, essentially, turning 
Eq. \ref{zj} around into a differential equation w.r.t. $\Lambda$.    
In general, for an action $S(\Lambda)$ of many fields $\phi_i$ valid for  a scale $\Lambda$, this equation will be of the form \cite{polch,wetterich}
\begin{equation}
\label{evolrenoaction}
\frac{dS}{d\Lambda}= \frac{1}{2} \mathrm{Tr}
 \left( \left( \frac{\partial^2 S(\Lambda)}{\partial \phi_i \partial \phi_j} +R_\Lambda\right)^{-1} \frac{\partial}{\partial \Lambda} R_\Lambda   \right)
\end{equation}
while correlation functions, $<k_1^i k_2^i ... k_i^i | k_1^f k_2^f ... k_j^f>$  , in turn, obey their local evolution equation which can be derived by differentiating Eq. \ref{zj}
\begin{equation}
\label{evolreno}
\left( \Lambda \frac{\partial}{\partial \Lambda} + \beta(\Lambda) +  \gamma   \right) <k_1^i k_2^i ... k_i^i | k_1^f k_2^f ... k_j^f>=0
\end{equation}
The parameters for these evolution equations are $R_\Lambda$, some smearing function encoding the renormalization scheme \cite{wetterich} (it must vanish at $k\ll \Lambda$ and diverge at $k\gg \Lambda$) and $\beta(\Lambda),\gamma,...$ are, respectively, the running of the coupling constant and scaling exponent.  The latter are related to the change in the effective action, and the number of degrees of freedom lost when an interval $d\Lambda$ is traced over.

What distinguishes renormalizable from non-renormalizable interactions is their asymptotic behavior as $\Lambda \rightarrow \infty$.
While renormalizeable interactions can exhibit a fixed point (which can be used to define an infrared theory independent of $\Lambda$), non-renormalizeable interactions (for example $\phi^{n>4}$) will typically come with factors of $k^n\Lambda^{-n}$ (this is clear from dimensional reasons), and hence will be invisible in the far infra-red.

If $\Lambda$ is a large but finite number (we expect the Planck energy $E_p$ to function as a physical $\Lambda$ for all quantum field theories), then, while renormalizeable terms dominate, we cannot automatically neglect non-renormalizeable ``irrelevant'' terms.
However, if the average momentum exchange $\ave{k} \ll \Lambda$, $\ave{k}/\Lambda$ continues to be a good expansion parameter which will allow us to define an ``effective Lagrangian'' to a desired order in $\Lambda$.
Hadronic theory with $\ave{k} \ll \lqcd$, and gravity at $\ave{k} \ll E_p $ can be thought of as effective theories of this kind. 

The distinction between renormalizeable and non-renormalizeable theories will play a secondary importance in the rest of this work, for we shall assume there is a physical cutoff scale, much higher than $\ave{k}$, distinguishing the ``fundamental'' and the ``effective'' theory.
\subsection{The density matrix in effective field theory \label{secdensmat}}
The above approach is useful if we have boundary conditions where the ``initial and final particle numbers'' are well-defined, and initial and final states can be prepared to be asymptotically well-defined.    An example are the textbook scattering problems.    More generally, due to energy conservation, any system where the total energy $\ll \Lambda$ can be expanded into a tractable sum of matrix coefficients  $<k_1^i k_2^i ... k_i^i | k_1^f k_2^f ... k_j^f>$, which can be calculated by systematically differentiating Eq. \ref{zj} (the shift $\order{\Lambda}$ will not matter in derivatives).

Let us, however, consider a system with a ``large number'' $N$ of particles, each with energy $k_i \ll \Lambda$, but where $\sum_i^N k_i \gg \Lambda$, and an experiment which cannot distinguish all particle final states with infinite precision (because, for example, it has no access to the required number of ``identically prepared systems'').

An example of such a system is a thermalizing QGP started by the collision of a large nucleus.    Another example, most likely, is a macroscopic black hole that forms and evaporates according to semiclassical emission of Hawking radiation.
Unless we consider a dilute system in a perturbative theory, sequential scattering can not be a priori assumed, and degrees of freedom are not on-shell momentum Eigenstates of a well-defined number of particles.   

In this regime, we cannot anymore automatically assume that the $k \ll \Lambda$ will produce a hierarchy of scales between the fundamental and the effective theory.  In the infinite volume thermodynamic limit $V\ave{k}^3 \rightarrow \infty$, indeed, we cannot automatically assume that the renormalizeable terms dominate over the irrelevant ones.

We can, however, still use
$<k_1^i k_2^i ... k_i^i | k_1^f k_2^f ... k_j^f>$ to construct a {\em density matrix} for the system and see how it evolves.   
Of course, we are talking about a Fock space, where every degree of freedom is bounded by the ``Microscopic scale'' $\Lambda$.   The total energy is however $N k \gg \Lambda$.  The density matrix is, therefore,
\begin{equation}
\label{densmat}
\hat{\rho} = \prod_{\otimes n,m } \prod_{\otimes p_1,p_2,...,p_n<\Lambda} \prod_{\otimes k_1,k_2,... k_m< \Lambda} \frac{1}{n!m!} c(k_1,...,k_n,p_1,...,p_m)^2 a^+(k_1)...a^+(k_n) |0><0| a (p_1)...a^+(p_n)    
\end{equation}
and the standard functional integral and, for a perturbative theory, Feynman diagrams techniques can be used to calculate the matrix elements $c(k_1,...,k_n,p_1,...,p_m)^2$ (they are related to the appropriate scattering matrix elements).
Note that the density matrix contains information about all the degrees of freedom of the system, {\em and their quantum correlations}, each of which has its own evolution equation of type Eq. \ref{evolreno}.   

Consider, furthermore, a situation where our experiment can only measure final state momentum Eigsenstates and S-matrices (all experiments discussed in section \ref{secdiscussion} are of this kind).     Because of the infinities pointed out in \cite{haag1,haag2} such detectors will therefore be very lousy density matrix measurers, because the number of correlations they can pick up is maximum $\sim \order{N}$ (for realistic experiments \cite{expcum1,expcum2,expcum3}, it is really $\order{1 \ll N}$, something we will elaborate in section \ref{secdiscussion}), and, for an EFT with a ``high'' cutoff $\Lambda \gg k$, they will miss all correlations where virtuality plays a part (as final state particles with sizeable virtuality corrections will be moving at near lightspeed).

For a  system with total energy $\gg \Lambda$, therefore,  even an ``ideal experiment'', capable of measuring correlations of an arbitrary high number of particles, will only measure as many $c(...)$ in Eq. \ref{densmat} as kinematics allows given the total system energy, an infinitesimal fraction of the possible ones.
``collective observables'', such as entropy\footnote{See footnote ``2'' at the beginning of the section \ref{secclassical}.  A ``state'' can likewise be interpreted in a ``frequentist'' and ``Bayesian'' way, and the density matrix is the quantum analogue of the phase space distribution function mentioned in section \ref{secclassical}} \cite{density}
\begin{equation}
\label{entropydef}
s =- \ave{\ln \hat{\rho}} =- \mathrm{Tr} \left( \hat{\rho} \ln \hat{\rho}\right) = -\sum_i   \lambda_i \ln \lambda_i
\end{equation}
where $\lambda_i$ are the probability coefficients for the system to be in each Eigenstate, will likewise be incompletely sampled, so even the entropy of a ``pure state'' will not be found to be zero.

Indeed, this problem has been known for a long time;   For instance, the discussion of Renyi entropy ($(1-l)^{-1}\ln\mathrm{Tr}(\rho^l)$, tending to Eq. \ref{entropydef} as $l\rightarrow 1$) measurements in \cite{bialas} parallels the issues explored, here, but the measurement in \cite{bialas} will not converge to Eq. \ref{entropydef} even for infinite multiplicity events sampled through infinite ensembles of identically prepared systems, where an $l \rightarrow 1$ limit is experimentally realistic.

This experimental ambiguity is complemented by a problem of describing $\hat{\rho}$ within effective theory: Since an effective field theory does not describe {\em every} correlation of the UV theory, it will not describe the entropy well.     In this case, even ideal experiments will not yield results that the ``effective theorist'' can interpret as a pure density matrix (carrying zero entropy), because, as shown in \cite{haag1,haag2}, the density matrices of the two theories are incommensurable.   

The ambiguity of the definition of entropy can be naturally understood from the fact that, in quantum mechanics, entropy conservation follows from unitarity \cite{density}.   For both normalizeable and non-renormalizeable theories, Eqs \ref{evolrenoaction} and \ref{evolreno} do not specify unitarity, which must be fixed, order by order in perturbation theory, via field strength renormalization \cite{peskin}.    We should not expect, therefore, Eq. \ref{entropydef} to be a conserved quantity for a theory where degrees of freedom are truncated and traced over.

For the ``large'' systems defined here, one can estimate the entropy an ``ideal experiment'' will measure by counting the $m$ elements of the density matrix in the accessible portion of the Fock space.  In a fully mixed theory,  Eq. \ref{densmat} will give an entropy of $m \ln m$.
As we will argue, as the number of correlations goes up the ability of effective field theories to describe them rapidly decreases.

We can make these considerations a bit more quantitative using the subadditivity constraints \cite{quantumentropy}:  If we denote the entropy of ``slow degrees of freedom'' by $S_{1}$ and the entropy of the ``fast degrees of freedom'' by $S_{2}$, then in general correlations between the slow degrees of freedom and the fast degrees of freedom account for a negentropy $S_{12}$, always bounded by
\begin{equation}
\label{ineq}
\left| S_{1} - S_{2} \right| \leq S_{12} \leq S_{1} + S_{2}
\end{equation}
Let us assume the ``fundamental'' theory is in a pure state, so $S_{1} + S_{2}=0$, but the correlations are hidden predominantly in the ``fast degrees of freedom'',
$S_{1} \simeq -S_{12},S_{2} \ll S_{1}$.    In this situation, a theorist with only the EFT at their disposal is bound to see a mixed state and an entropy above $0$. 
Looking at Eq. \ref{ineq}, with $S_{2}$ representing the entropy contained in a segment $d \Lambda$ shows that entropy cannot be renormalized via an equation such as Eq. \ref{evolreno}:     If the fundamental theory has zero entropy, the entropy traced over cannot be considered ``small''.    In a strongly coupled theory, where correlations are strong, it is not immediately obvious that the entropy contained within $S_{2} \ll S_{1}$ even if $d\Lambda \ll \Lambda$.

To see when the situation described here is likely to occur, consider an effective theory defined on a lattice in a box of length $L$, with a cutoff scale $\Lambda$, so that for simplicity $\Lambda L = 1/m$ where $m$ is an integer $\gg 1$.  For simplicity we consider only one degree of freedom, as in the Lagrangian of Eq. \ref{scalarfield}.  This is actually a good effective theory for a gas of pions close to ``freezeout'' \cite{jansbook}, at densities $\ll \lqcd^3$.
  
If the effective theory appears ``free'' (i.e. is weakly coupled) particles with different occupancy numbers $i=1 ... m$ are distinguishable.
Truncating modes at $k \leq \Lambda$, and enforcing energy conservation, we recover a very large but finite-dimensional Hilbert space.
The density matrix of such a system will be of the form
\begin{equation}
\label{rhofree}
\hat{\rho}_{free-IR} \sim \sum_{\sum j\times i = m,\forall i \ll m} \alpha_{i j} a^+(i)^j |0><0|a^j(i)
\end{equation}
Such a matrix will have $\sim n^\zeta$ Eigenvalues, where $\zeta$, for bosons, is some number set by total kinematics and $0,1$ for fermions.  Each term will, according to the Equation \ref{entropydef}, contribute from zero to $\sim \alpha \ln n$ to the entropy.  By construction, $n\ll m$.

This will also be what a density matrix of a theory interacting in the UV and free in the IR {\em as calculated by the effective IR free theory} will record.  At least kinematically, such interactions are allowed for all $m\geq 1$.
The ``true'' density matrix, beyond the EFT, of the pure state and containing all  interactions will however be of the general form
\begin{equation}
\label{rhoint}
\hat{\rho}_{interacting-UV} \sim \sum_{i\times j=m} \sum_{k \times l = m}  \alpha_{ijkl} (a^+(i))^j |0><0|(a^+(k))^l
\end{equation}
this expression has, potentially, $m^\alpha \gg n^\alpha$ Eigenvalues, each contributing as $\sim \alpha \ln \alpha$.   Furthermore, in a basis where  Eq. \ref{rhofree} is diagonal, Eq. \ref{rhoint} will not be, since interactions mix terms of different momentum ($i,k$ in Eq. \ref{rhoint}).   It is clear that as $\Lambda \rightarrow \infty$, the matrix \ref{rhoint} traced over the fast degrees of freedom will become more and more orthogonal to the matrix in \ref{rhofree}, and indeed, this is one way to see the theorems in \cite{haag1,haag2}.

A theorist with only a free theory at their disposal will continue to neglect the $(m^n)^\alpha-n^\alpha$ terms separating Eq. \ref{rhoint} from Eq. \ref{rhofree}.For certain calculations this is indeed acceptable:  
If we consider $e^+,e^-$ and $pp$ collisions testing the standard model, provided this model is stable against some scale much higher than the Electroweak symmetry breaking, the ``box'' on which our ansatz is based on is limited, kinematically, to $m\leq 1$.   

But for black holes of mass much bigger than the Planck length, or hadron collisions with $\sqrt{s} \gg \lqcd$, $m\gg 1$ and any correlation $i \times k \in 1,...,m$ in Eq. \ref{rhoint} is likely to carry an $S_{12}$ negentropy term.
However, ``Entropy'' in a hadronic collision has been historically \cite{Fer50,jansbook} measured by counting the only ``free'' final state pions in an ``average'' event, while in black holes thermal formulae of a free theory in a curved background are usually used.    
The crucial question is, how important are the interactions not captured by the effective theory, i.e. what fraction of the $(m)^\alpha-n^\alpha$ terms are actually relevant?

If the fundamental theory at $\Lambda$ is strongly coupled, we can reasonably suppose that the classical analogue of the system at $\Lambda$ is chaotic. We can then further our insight by applying the Berry conjecture \cite{berry1}, stating
{\em wavefunctions of high-lying energy levels of classically chaotic systems are pseudo-random functions}.   While this conjecture has never been extended beyond quantum mechanics, we feel confident using it because of the truncation in $\Lambda$.
In this case \cite{berry2,berry3}, the coefficients in Eq. \ref{rhoint} will be random matrix elements, constrained by the global unitarity of $\hat{\rho}$ ($\mathrm{Tr}\hat{\rho}_{interacting-UV}=1$).
In this limit, all allowed $n\leq m$ entanglements are equally likely.
Hence, we can confirm that the negentropy $J_N$ of the quantum state
\begin{equation}
J_N = s_{max} -s = s_{thermal}-s \simeq \order{(L\Lambda)^3}-s 
\end{equation}
where $L$ is the system size, will be dominated by $\order m^2$ $S_{12}$ terms correlating {\em all} the degrees of freedom of the system.

In other words, if only the terms in Eq. \ref{rhofree} are sampled, one will see an entropy $\gg 0$ because the state looks mixed in the IR, and the negative terms correlating the IR and UV degrees of freedom will not be ``seen'' within effective theory.   Such terms will also only be measured by an experimentalist able to measure correlations of the order of the total number of particles in the system.

A more direct, but less physically intuitive proof that entropy is generally specific to the timescale can be obtained by considering quantum discord \cite{discord}, 
\begin{equation}
D_{IR,UV} = s_{IR} - s_{IR,UV} + s_{IR|UV}
\end{equation}
where
\begin{eqnarray}
s_{IR,UV}&=&   - \mathrm{Tr}_{IR,UV} \ln \rho^{IR,UV} \ln(\rho^{IR,UV})\phantom{AA}\\  s_{IR|UV}&=& -\mathrm{Tr}_{IR} \ln \left( \rho^{IR} \mathrm{Tr}_{UV} \rho^{IR,UV} \ln\left(\rho^{IR,UV} \right) \right)
\end{eqnarray}
it is explicitly clear that for a theory where interactions entangle UV and IR degrees of freedom (ie, any theory with interactions mixing UV and IR) the quantum discord will depend on the scale separating UV and IR.   Which means that entropy will generally depend on $\Lambda$.   The condition for interactions mixing IR and UV degrees of freedom ``in the infrared'' is simply that such interactions are kinematically allowed, i.e. that the total size of the system is of the order as the ultraviolet scale.

The above reasoning is based on the density matrix, and hence is valid independently from whether the system in question starts out in a pure state with zero entropy, or is close to equilibrium.   In the next section, we shall explore, on a qualitative level, how the $IR$ density matrix is expected to evolve in time from an initially pure state. 
\subsection{Dynamics of the density matrix in an effective field theory \label{secdyn}}
\begin{figure}[h]
\begin{center}
\epsfig{width=18cm,figure=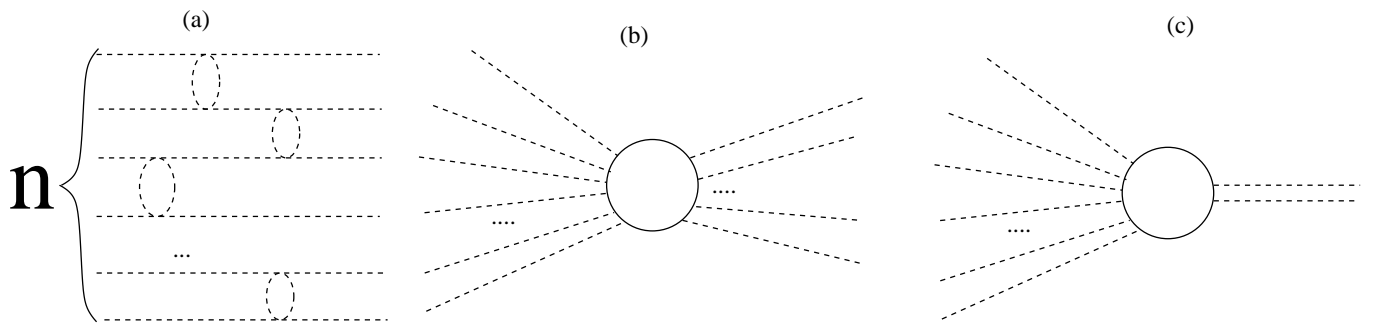}
\caption{\label{eftcoarse} A fully renormalized diagram (panel (a)), a diagram capable of coupling pure and mixed EFT states (panel (b)) and a diagram capable of moving energy-momentum from the slow to the fast degrees of freedom (panel(c)).
Here, dashed lines represent IR $\phi$ fields, while solid loops give traced-over UV degrees of freedom. 
}
\end{center}
\end{figure}

Going further with a quantitative calculation in such a general setup is, of course, extremely complicated.
It is not the purpose of these notes to do this, but just to see what happens if we try to ``renormalize'' $\hat{\rho}$, in other words to try to see whether we can hide the ``missing terms'' in a $\Lambda$-dependent redefinition of $\hat{\rho}(\Lambda)$.
We immediately see (Fig. \ref{eftcoarse}, using the Lagrangian of Eq. \ref{scalarfield}) that terms where $\sum_i^n k_{i}\sim \Lambda$ will become a problem for any theory with a cutoff, renormalizeable or not, because $n-particle$ correlations mix slow and fast degrees of freedom.
Again, because $\hat{\rho}$ counts {\em all} kinematically allowed  states in Fock space, one has to consider diagrams of an arbitrarily large number of in and out states.  As one goes higher in in and out states, the sensitivity to ``irrelevant'' degrees of freedom increases.

Consider the Feynman diagram expansion for the weakly coupled $\phi^4$ theory described in the previous section \ref{secrenoreview},
a diagram such as (a) of Fig. \ref{eftcoarse} is constructed entirely of ``slow'' in and out states.    It will therefore lead to a unitary evolution in the truncated theory only.    Any loops and infinities will be canceled out by suitable counterterms, arising from diagrams (b)

The diagram (c), {\em of the same order} will however correlate slow and fast degrees of freedom.    In a Wilsonian RG flow, such diagrams will be traced over, and any information coming from them is undetectable in the purely slow theory.
Correlations, in the UV theory, resulting from such diagrams will be invisible in a density matrix $\hat{\rho}_{IR}$ defined at scales $\ll \Lambda$, and any conservation law appearing in the IR (unitarity, but also energy-momentum conservation) can be violated.  
  Physically, such quantum fluctuations transforming ``n slow fields into a fast field'' will show up as purely imaginary terms into an IR-effective lagrangian.

While these diagrams are generally suppressed as $f(k)^n \Lambda^{-1}$ consider, panel (b) of Fig. \ref{eftcoarse}, are not, since in these diagrams fast degrees of freedom appear as virtual states only.
Such processes will respect energy conservation but can cause transition between a pure and a mixed state of the EFT.
Hence, as also clear by the optical theorem, the effective term showing up in the IR Lagrangian from such terms will be complex.
Such terms are of course suppressed by $k/\Lambda$, but are correspondingly enhanced as $\sim n!$ number of possible diagrams.

Summarizing, an observer living at a scale $\ll \Lambda$ where they use an effective theory characterized by Eigenvalues $|i_E>$ of energy $E$ will see the dynamics of a system of size $\gg \Lambda$ as being described by a {\em non-unitary} effective Lagrangian
\begin{equation}
L_{eff}^n = \ave{M}_{(a)}^2 \phi^n_{slow} + f(n)\frac{1}{\Lambda^n} \left(i \ave{M_c}^2 \phi^{n/2}_{slow} + \ave{M_b}^2 e^{b+i c} \phi^n_{slow} \right)
\end{equation}
transitions of the type corresponding to the last two terms are forbidden, respectively, by conservation laws ($M_b$ terms allow for transition Eigenstate having different energy) and unitarity ($M_c$ terms allow for transitions between pure and mixed Eigenstates) .  Indeed, within an eventual Fock space of the ``fundamental theory'' (QCD for heavy ions, full quantum gravity for black holes, and so on) such terms will be real again because states with $E \geq \Lambda$ will be included in this theory's Fock space.

In a scattering problem, such terms are irrelevant because we {\em assumed} $\sqrt{s} \ll \Lambda$.   But for systems whose {\em total energy content } $\gg \Lambda$, even for ``dilute systems'', we cannot automatically make such an assumption.    What this means is that we cannot expect the evolution of any effective $\hat{\rho}(\Lambda)$ to be unitary even when the average momentum per degree of freedom $k \ll \Lambda$.
Furthermore, while complex effects on n-particle correlations are suppressed by $(k/\Lambda)^n$ they are effectively enhanced by the $n!$ number of such diagrams.   For any finite $\Lambda$, there will be a point where $(\Lambda/k)^n \sim n!$.   At this point, the infrared EFT cannot reliably be used to compute such $n$ particle correlations.   For average quantities, such as the occupation number, such terms can be reasonably assumed to be highly suppressed.   For $n-$particle correlations, however, such terms are {\em leading}, and the negentropy of the system contains correlations of arbitrary $n$.

Hence, the negentropy in such traced-over correlations is missing from the ``effective entropy'' operator $-\ave{\ln \hat{\rho}(\Lambda)}$, which therefore will not be conserved.   
In particular, a pure state, where $\rho = \delta_{i 
m}(\Lambda)$ where $m$ is {\em some} state in Fock space will appear as a mixed state both after some evolution time and by just changing $\Lambda$.

\begin{figure}[h]
\begin{center}
\epsfig{width=10cm,figure=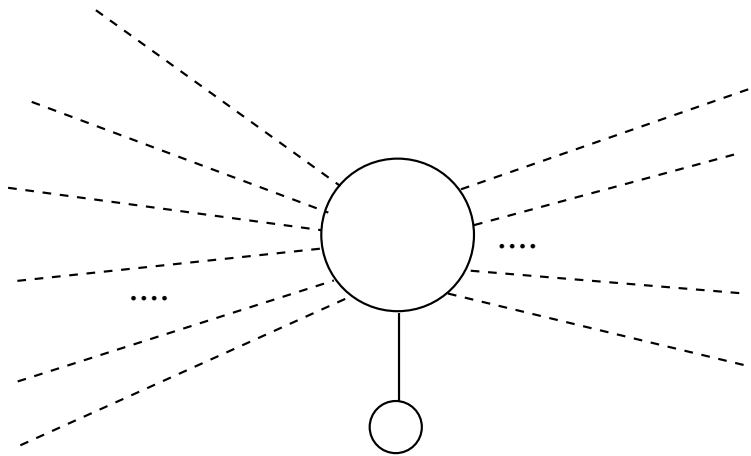}
\caption{\label{eftcoarse2} UV effects on the ``counterterm'' of a scattering which appears to be purely in the IR }
\end{center}
\end{figure}

The Feynman diagram expansion suggests that n-th order diagrams, controlling n-particle correlations, are the full quantum equivalent of n-th order BBGKY hierarchy equations.   ``Truncating at the nth order'' as per Eq. \ref{truncs} is equivalent to assuming that, at that order, traced over ``fast'' degrees of freedom are dominant, and all correlations disappear.
We shall now attempt to justify this suggestion, and estimate, as a function of $k/\Lambda$, the timescale for such correlations to disappear.   To do this, we try to understand how the density matrix evolution equation
\begin{equation}
\frac{d\hat{\rho}}{dt} = i \left[ \hat{H},\hat{\rho} \right] \Rightarrow \hat{\rho}(t) = \exp\left[ -\hat{H}(t) \right]\hat{\rho}(0) \exp\left[ \hat{H}(t) \right] 
\end{equation}
changes when $\hat{\rho}(\Lambda)$ is the density matrix of a field theory coarse-grained to an effective theory.   We shall denote as $\hat{\rho}_{slow}$ terms in Eq. \ref{densmat} where $\forall k_i,p_i \leq \Lambda$ and $\hat{\rho}_{fast}$ all other terms.
We shall also denote as $\hat{H}_{slow-slow}$,$\hat{H}_{slow-fast-slow},\hat{H}_{slow-fast}$ as the Hamiltonian terms resulting from diagrams such as those of Fig. \ref{eftcoarse}  (a),(b),(c).   We note that the latter two, in general depend both on slow and fast degrees of freedom.
The time evolution of the coarse-grained density matrix becomes
\begin{equation}
\label{rhoslow1}
\frac{d \hat{\rho}_{slow}}{dt} =i \mathrm{Tr}_{\phi_{fast}} \left[ \hat{H}_{slow-slow} + \hat{H}_{slow-fast-slow} + \hat{H}_{slow-fast},\hat{\rho}_{slow} \otimes \hat{\rho}_{fast}     \right] 
\end{equation}
Furthermore, since correlations are traced over, the diagrams with an intermediate fast stage in $H_{slow-fast-slow}$ are generally non-linear, since the contribution of diagrams such as (c) in Fig. \ref{eftcoarse} as well as Fig. \ref{eftcoarse2} to occur depend on the amplitude of the ``slow'' incoming states.    
Finally, we note that energy is not conserved for the slow degrees of freedom at anyone time, since part of the energy goes, on average into the traced-over fast degrees of freedom.   We shall denote as $E',E-E'$ as the energies present in the slow and the fast degrees of freedom, and label $\hat{\rho}_{slow,fast}(E)$ accordingly.

Integrating out the fast fields the equation \ref{rhoslow1} will therefore look something like
\begin{equation}
\label{evolrho2}
\hat{\rho}_{slow}(E,t) = \int dE' dt \exp\left[-i \hat{H}_{eff}\right]\hat{\rho}_{slow}(E,0) \exp\left[ -i \hat{H}_{eff}\right]
\end{equation}
where $H_{eff}$ can be decomposed into a perturbative series whose coefficients $\hat{H}^n$ will generally be of the form
\begin{equation}
\label{evolrho}
 \hat{H}_{eff} =    \hat{H}_{slow-slow}\left( E' \right) + \sum_{n=1}^\infty \left( \frac{k}{\Lambda} \right)^n \times
\end{equation}
 \[\  \times \left( \ave{\hat{H}^n_{slow-fast-slow} \left( E',\hat{\rho}_{slow}(E') \right)}_{fast}
+  i\ave{\hat{H}_{slow-fast}^n \left( E,E',\hat{\rho}_{slow}(E') \right)}_{fast} \right)
\]
and $\ave{...}_{fast}$ means traced over fast degrees of freedom.
this is, generally, a non-linear dissipative evolution equation, with the non-linear and dissipative part suppressed by a power series in  $(\sim k/\Lambda)^n$.
The corresponding Hamiltonians $\hat{H}^n$ can be obtained by counting all possible Feynman diagrams of the appropriate type.   As can be seen by comparing Eqs \ref{rhofree} and \ref{rhoint}, this means correlations of $\sim N \sim E/k $ particles will eventually be dominant in the density matrix.
At this point, even if a projection to an $n-$particle Hilbert state carried virtually all of the entropy initially, i.e. the Density matrix of Eq. \ref{densmat} reduces to 
\[\ 
c(k_1,...,k_i,p_1,...,p_j)^2 \simeq \delta_{i,n}\delta_{j,m} \prod_i \delta_{k_1,K_1}
 \prod_j \delta_{p_i,P_k} \]
for a particular set of $\{ K,P\}$, after a time scale $\sim \tau \left( \Lambda/k^2 \right)$ {\em all} elements of the density matrix will be equally represented
, and the infrared theory will miss most of the entropy.

The evolution of Eqs \ref{evolrho},\ref{evolrho2} includes both a small dissipative term and a larger non-linear term.   Such systems typically admit both chaotic and deterministic regimes \cite{classical1,classical2}, depending on the relative importance of the two terms.   Which is the case will determine the form of $\tau(...)$.

If Eq. \ref{evolrho} is chaotic, as can be expected for a strongly interacting system at $\Lambda$ (both gravity at the Planck scale and QCD at the scale $\sim \lqcd$ are in this class), then even if the Lyapunov exponent $\sim \ave{k}/\Lambda$ after a time
$\sim \ln \left( \Lambda/\ave{k}  \right)/\ave{k}$ effective field theories will be a lousy description of the density matrix in the infrared.    Ever for a large separation of scales, this might result in a parametrically small amount of time.     A good heuristic model of such a situation is the Kolmogorov cascade \cite{lifs}:  High occupancy semiclassically evolving soft modes become unstable, cascading into higher frequency lower occupancy modes.  Systems exhibiting this behavior, studied in the classical and semiclassical limit  \cite{weibel,mrow,dusling} typically show a parametrically fast ``isotropization'' via cascading instabilities, something that is often claimed ``fakes'' thermalization.
In the classical picture, thermalization is indeed fake because the system continues to be in a coherent state by construction.
However ,once higher order quantum correlations are included, and fast degrees of freedom are traced over, ``fake'' thermalization will be indistinguishable from a real maximally mixed state.    This suggests that,
for any cutoff scale separating relevant from irrelevant degrees of freedom, even a weakly coupled effective description can lose significant amounts of unitarity in parametrically short timescales.

If Eq. \ref{evolrho} is not chaotic, the timescale for the same process to happen is  $\sim \Lambda/\ave{k}^2$, which goes to infinity in the IR limit.   Even in this case, however, a macroscopic system which is also comparatively long-lived in a dense phase could functionally also look ``thermalized'' after it decouples.
 
In the next two subsections, we shall look at two examples of current theoretical and phenomenological interest where additional scales can ``hide'' the effects discussed above:  Theories with broken continuus symmetries, admitting Goldstone eccitations, and theories in the planar limit.
\subsection{ Theories with broken continuus symmetries \label{secgoldstone}}
The discussion above has to contend with a seemingly obvious counter-example:  Condensed matter physics is rich with examples of ``large'' systems created in the laboratory where effective degrees of freedom evolve according to unitary theories, and no dissipation between ``slow'' and ``fast'' degrees of freedom is observed:  If one considers the total energy of the system, processes breaking up phonons in superfluids and magnons in magnets are abundantly allowed kinematically.    Yet, propagation of these degrees of freedom in a material can be described by a unitary theory to an excellent approximation \cite{zee}, and very little interaction between traced-over and effective degrees of freedom of the type described in the previous section, is observed.

The condensed-matter systems in question, however, have to important differences w.r.t. the generic setups described in the previous section:
The first difference is the fact that such systems are very close to the ground state, with $\mu \gg T$ and most of the relativistic energy stored in mass density rather than in excitations.   As Berry's conjecture only applies to ``highly excited'' states \cite{berry1,berry2,berry3}, where effects of the energy gap are negligible, we cannot assume the density matrices of such systems are ``random'' rather than dominated by low-level excitations.
If, in addition to the vacuum, one has excited only ``a few'' degrees of freedom, and one keeps track of both incoming and outgoing Goldstones, then the energy gaps (as well as the Fermi surface, in the Fermionic system) protects the unitarity of the excited degrees of freedom's evolution.   In this case, the situation is similar to the ``elementary scattering'' with $\sqrt{s} \ll \Lambda$, where no ``extensive scale'' $\gg \Lambda$ exists and the density matrix only has a few elements, all ``slow''.
What happens, however, when ``many'' phonons or magnons, forming a ``large system'' of total energy $\gg \Lambda$ (which, for a condensed matter system, might be the inverse of the molecular size) are excited?
\begin{figure}[h]
\begin{center}
\epsfig{width=18cm,figure=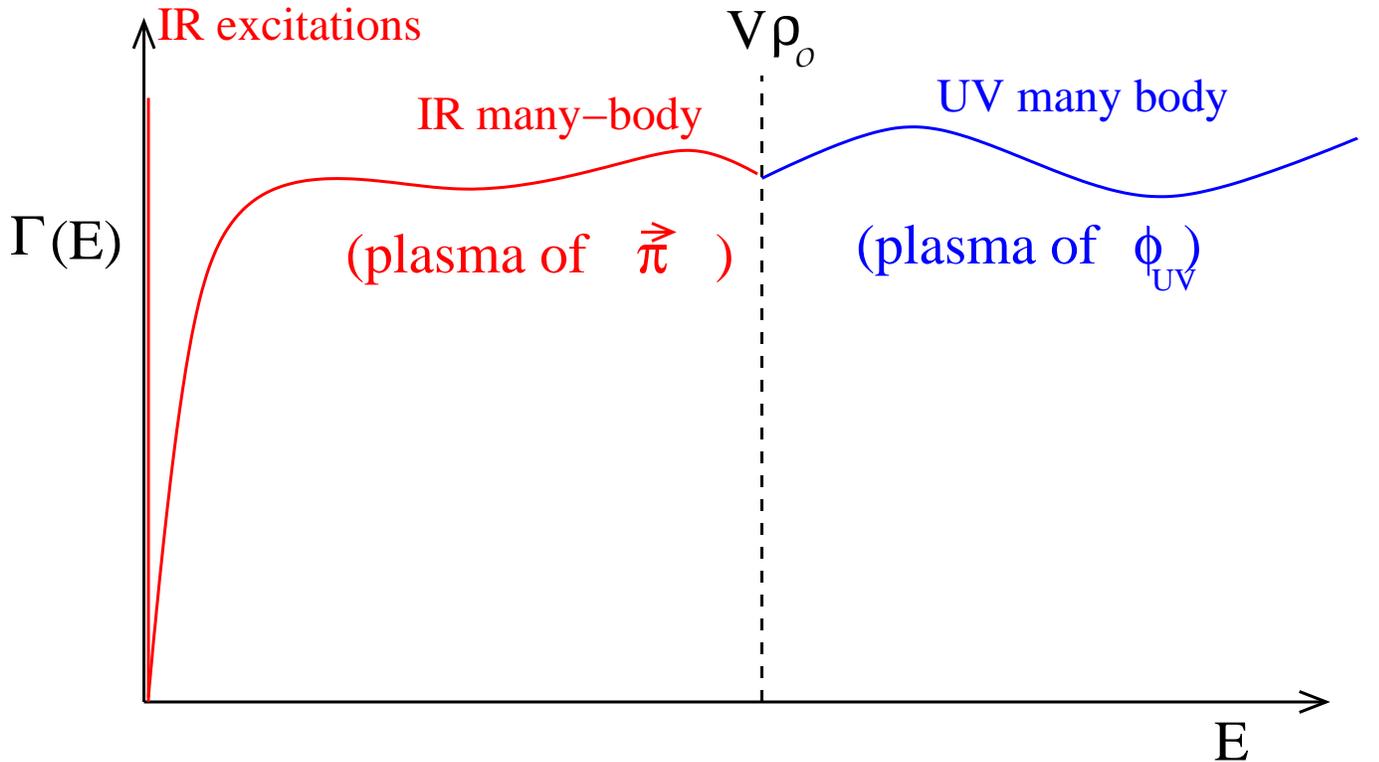}
\caption{\label{goldstone} The spectral function of a ``large'' system whose microscopic dynamics exhibits broken symmetry, with a condensate $\ave{\mathcal{O}}$ and Goldstone eccitations $\vec{\pi}$ and ``fundamental'' degrees of fredom $\phi_{UV}$.
}
\end{center}
\end{figure}

In this case, it must also be kept in mind that the typical condensed matter system described well by such an effective quantum theory \cite{zee} is characterized by the presence of a condensate (generically denoted here by $\ave{\mathcal{O}}$, spontaneusly breaking a symmetry of the system (respectively, the global phase of the condensate in a superfluid, electromagnetic $U(1)$ for cooper pairs and rotational invariance for magnetized materials).   The effective theory which works so well is, indeed, a theory of Goldstone bosons (phonons and magnons) for the broken symmetry.

In this case, the basic hyerarchy of scales described in section \ref{secdensmat} becomes more complicated, as the condensate enters as an additional scale, both intensive and extensive, to be dealth with.
To see this, let us consider a theory with a condensate $\mathcal{O}$, carrying energy density $\rho_\mathcal{O}$ and admitting Goldstone excitations $\vec{\pi}_\mathcal{O}$.   Let us, as in sections \ref{secdensmat} and \ref{secdyn} put this sytem in a volume $V$, and energy density $\rho$ (mean momentum exchange $k \sim \rho^{1/4}$).   The spectral density of such a generic theory is shown in Fig. \ref{goldstone}.

While it is tempting to treat $\rho_\mathcal{O}^{1/3}$ as the $\Lambda$ discussed in sections \ref{secdensmat} and \ref{secdyn}, destroying the condensate over a large volume $V$ requires an energy $\sim \rho_\mathcal{O} V$.   For a cold and dilute system, where $\rho \ll \rho_{\mathcal{O}}$ the ``slow-fast'' interactions discussed in \ref{secdyn} cannot, kinematically, destroy the condensate in the macroscopic region.   Hence, the whole density matrix $\hat{\rho}$ can be effectively built out of the effective degrees of freedom $\vec{\pi}_\mathcal{O}$.

If we only see a subset of excitations of an entire system up to an energy $\Lambda$, so a hyerarchy $\Lambda \ll \rho V \ll \rho_\mathcal{O} V$ holds, we might well see a ``plasma of Goldstone bosons'' (a liquid of phonons within a superfluid), thermalizing within the condensate in a timescale as a function of $\Lambda/k$, as described by section \ref{secdyn}.    In this case, if the Goldstones are ``strongly interacting enough'', the coefficients of the density matrix $\hat{\rho}$ will indeed be close to random, but the basis vectors characterizing the density matrix will be Eigenstates of the effective Hamiltonian, as the condensate will ensure the ``UV'' degrees of freedom are unoccupied.

To summarize, the density matrices in different regimes will be
\begin{equation}
\hat{\rho}_{k \ll \Lambda \ll \rho_\mathcal{O} V} = \alpha^{\mathrm{IR}}_{k k' \ll \Lambda} \left| \vec{\pi}_\mathcal{O}(k) \right>    \left< \vec{\pi}_\mathcal{O}(k') \right| \eqcomma    \hat{\rho}_{k \ll \Lambda \gg \rho_\mathcal{O} V} = \alpha_{k k' \ll \Lambda}^{\mathrm{UV}} \left| \phi_{\mathrm{UV}}(k) \right>    \left< \phi_{\mathrm{UV}}(k') \right|
\end{equation}
where the $\phi_{\mathrm{UV}}$ are the UV symmetry-restored degrees of freedom. As long as $\Lambda \ll \rho_\mathcal{O} V$, mixings between symmetry-broken and symmetry restored states are kinematically not allowed, and hence the IR theory will appear unitary.   The ``tracing over'' of Goldstone degrees of freedom $\ll k$, and randomization of $\alpha^{\mathrm{IR}}$ through reactions such as those in section \ref{secdyn} is however possible.    Thus, while it is plausible that bringing $\Lambda$ close to $\rho_\mathcal{O} V$ will make the theory more strongly interacting as the symmetry is being restored, thermalization should not uniquely signal a change in degrees of freedom within the system.

\subsection{Theories in the ``planar limit'' and Gauge/Gravity duality \label{secplanar}}
Looking at Eqs \ref{rhoint} and \ref{rhofree}, one can see that 
an alternative way to improve the performance of EFT in entropy calculation, without raising $\Lambda$, is to increase the microscopically distinguishable number of degrees of freedom, while, at the same time, decreasing the probability of two distinguishable degrees of freedom entangling.   In this limit, the entropy content in Eq. \ref{rhofree} $\rightarrow \infty$ ( occupancy number $\rightarrow 0$, so the $-\ln(...)$  term in Eq. \ref{entropydef} $\rightarrow \infty$  ) while the nonlinear and dissipative terms in  Eq. \ref{evolrho}, decreases in amplitude by a parametric amount.    This is exactly what happens in the large $N_c$ expansion \cite{largenc}, where a the dimension of the Gauge group $SU(N_c \rightarrow \infty)$.  In this limit the number of colors $N_c$ and the coupling constant $g_{YM}$ increase and decrease respectively, maintaining a constant $g_{YM}^2N_c$. This expansion persists in the Gauge/Gravity duality picture \cite{adscft1,adscft2,adscft3,adscft4} where,in fact, $N_c \rightarrow \infty$ ``faster'' than $\lambda \rightarrow \infty$.
Thus, we expect that the effects described here cannot be captured in ``planar expansions'', where the classical approximation continues to describe the evolution of the density function well.   

In addition, specifically for the AdS/CFT correspondence \cite{adscft1,adscft2,adscft3,adscft4} it is worth noting that in AdS space the $\Lambda/k^2$ scale invoked previously for freezeout is actually infinite: black holes in AdS are stable \cite{hawkingpage}, so, by the arguments described here, density matrices in a black hole background are ``as random as they can be'' for any value of the coupling. In the dual picture, the effect of running of any renormalizeable parameters vanishes (terms such as Fig. \ref{eftcoarse} panel (a) are scale-invariant), while, as usual, irrelevant parameters vanish for $\Lambda \rightarrow \infty$.   In both of these cases, therefore, the considerations of this paper are most likely irrelevant, because parameters we assumed to be finite in this work diverge.  Since these parameters are finite in the real world, however, this suggests that ``real world physics in the IR'' is radically different from that captured in \cite{adscft1,adscft2,adscft3,adscft4} as far as thermalization is concerned.

Strong hints that the planar limit $\mathcal{N}=4$ SYM is integrable \cite{integ}, yet, unlike classical integrable systems \cite{classical1,classical2}, exhibits a finite viscosity\cite{viscosity}\footnote{It should however be kept in mind that the definition of integrability used in most of the works of this kind, in terms of S-matrix factorization, somewhat differs from the most direct classical analogue:  The $S$-matrix does not capture {\em all} dynamics of operators, but only its asymptotic part.  It is therefore not so clear that, analogously to classical systems, ``integrable'' quantum systems do not thermalize;  Semiclassically, the molecular chaos assumption is only valid at distances parametrically larger than the microscopic scale \cite{dirk}  } and thermalization time \cite{chesler}, reinforce this supposition.
It is at first sight puzzling that an ``integrable'' zero temperature theory can thermalize, the process dual to the formation of a black hole from an initial distribution in AdS space of a large (w.r.t. the Planck scale) amount of matter.
After this matter crossed the event horizon, however, back-reaction terms and quantum corrections to gravity, dual to $N_c^{-1}$ corrections in the CFT, cannot anymore be neglected.    This apparent paradox is a good illustration that, while the planar limit can be used to ``hide'' the negentropy due to interactions, this negentropy cannot be neglected when calculating the total entropy of this system.  

To summarize this discussion, care must be taken when calculating correlations between ``many''
 degrees of freedom in a ``many particle system'' using an EFT.
If the energy of the ``total system'' is comparable to the expected cutoff scale, even in a renormalizeable theory ``irrelevant'' operators will show up when many-particle correlations are considered.   When entropy is considered within an EFT, these operators will manifest themselves as a loss of unitarity, to reflect the flow of information across the cutoff.  In the concluding section we examine possible ``phenomenological'' consequences of these effects.
\section{Discussion \label{secdiscussion}}
Let us consider a ``scattering'' where a strongly interacting and dense system (intensive scale $\sim \Lambda$, with a global energy $\gg \Lambda$) is produced from a pure state,   but then expands and decays into a ``many''-particle system of typical momenta $k \ll \Lambda$ (hence, there will be $N \sim \Lambda/k\gg 1$ particles at the end). A good example of this is an expanding and hadronizing Quark-Gluon plasma, but any ``large bag of particles'' will do.   Crucially, a black hole of mass $\gg 10^{19}$ GeV formed from a ``pure'' quantum state and evaporating should also fall into this category.
The crucial question is: to what extent can an EFT, valid for $\ll \Lambda$, be used to describe the final state of the system.

As the previous sections argued, if the system is strongly interacting (as in quantum gravity at momentum scales $\sim E_p \sim 10^{19}$ GeV, or QCD with momentum scales $ \lqcd\sim 300 $ MeV), or even a weakly coupled theory where the typical state occupancy number $N_{state} \gg 1$ (of the type studied in \cite{schwinger,weibel,mrow,dusling}), it is natural to assume that en evolution described by Eq. \ref{evolrho},\ref{evolrho2} will result in very complicated correlations of a large ($\sim N$) numbers of particles.

If an ``ideal'' detector accepts all particles produced in the event, {\em and } enough ``identically prepared systems'' are available\footnote{See footnote ``2'' at the beginning of section \ref{secclassical}} to measure correlations of {\em all} produced particles, than, provided the black hole or fireball was produced within a pure state, the experiment could measure only a ``modest'' increase in entropy, reflecting information lost via the coarse-graining barrier (Panel (c) in Fig. \ref{eftcoarse}).   Such reactions are kinematically allowed, but strongly suppressed, since they require $k \sim \Lambda \gg \ave{k}$.

If,however, such a precision is not available, then the ``realistic detector'' will find that the original pure state produced a large amount of entropy in its evolution: The number of particles most likely increased, as the initial kinetic energy went into particle production, and the correlations which balance the increased number of degrees of freedom ($\sim N^N$ negative entropy $S_{12}$ terms in Eq. \ref{ineq}) are invisible to the detector.

This increase in entropy will be interpreted by the ``effective theorist'', with only the effective field theory at their disposal, as a violation of unitarity, for such a theorist can only calculate correlations of up to $\ll N$ particles reliably.   The ``fundamental theorist'', with access to the full strongly interacting fundamental theory will be able to predict, from fundamental theory and particle averages, the necessary $N-$particle correlations which will, to a good approximation, restore unitarity.   An ``ideal detector'', capable of measuring correlations between $\sim N$ produced particles, will however be necessary to verify these predictions.   

The near-thermal particle distributions observed so far in {\em all} hadronic collisions are therefore natural, for realistic experiments are very far away from the ``ideal detector'' limit described above.     It is therefore not surprising that, as known for a long time, thermal model describe very well a closed quantum system such as a hadronic collision, even for systems much smaller than those where kinetic thermalization is required ~\cite{Fer50,Pom51,Lan53,Hag65,pbm,jansbook,becattini}, as long as the observables considered are final-state particle averages. The distinction between ``fake thermalization'' and ``real thermalization'', if such a thing is meaningful, however, requires measuring higher and higher particle momentum distribution cumulants and comparing them to statistical mechanics expectations.   We do not know how realistic is such a program. Measurements of this type have been performed for other reasons in the past \cite{expcum1,expcum2,expcum3}, but for obvious reasons they fall far short of the ``ideal detector'' standard described above.

Note that to reach the conclusions above we just needed to assume that the final state is described by an EFT (the typical momentum is $\ll \Lambda$), but the system itself is ``large'' w.r.t. $\Lambda$ and exhibits a ``strongly interacting'' phase.
The above discussion, from an EFT point of view, is therefore similar to the ``scattering'' of a pure quantum system into a black hole, which eventually evaporates into Hawking radiation: 
General relativity, coupled to quantum field theory, can be considered as an effective field theory to a hitherto unknown UV-complete quantum theory of gravity, which is relevant for momentum exchanges of order $E_p$ GeV.

The black hole evaporation process, similar to its sonic counterpart \cite{unruhole} and the Schwinger pair production mechanism in a strong field \cite{schwinger}, is a semiclassical but non-perturbative process involving an ``infinite'' ($N \gg 1$) number of ingoing and outgoing states.    Infinite, here, simply means that backreaction of evaporated particles on the parent system can safely be neglected.   Semiclassical means that quantum fluctuations around the action extremum can be neglected.   Quantitatively, the two assumptions are connected since such corrections will carry $\order{1}$ ``units of action'' ($1$ in natural units, $\hbar$ in other unit systems) while the system as a whole carries $\order{N} \gg 1$ units.

This picture is intuitively clear in the laboratory ``analogue of Hawking radiation'', the sonic black hole \cite{unruhole};   In this case, the physical picture, and the analogy with \cite{schwinger}, is clearly that of a mean-field condensate amenable to a coarse-grained description (the ``fluid'') and an effective quantum theory of perturbations (the ``phonons'').    The unitarity violation implicit ``Hawking sound'' is clearly of the same type as that of \cite{schwinger}, and can be accounted for by going further than the semiclassical approach.
To what extent a ``real black hole'' is analogous to its sonic counterpart of course remains to be seen, but certainly this scenario is the simplest one.

\begin{figure}[h]
\begin{center}
\epsfig{width=18cm,figure=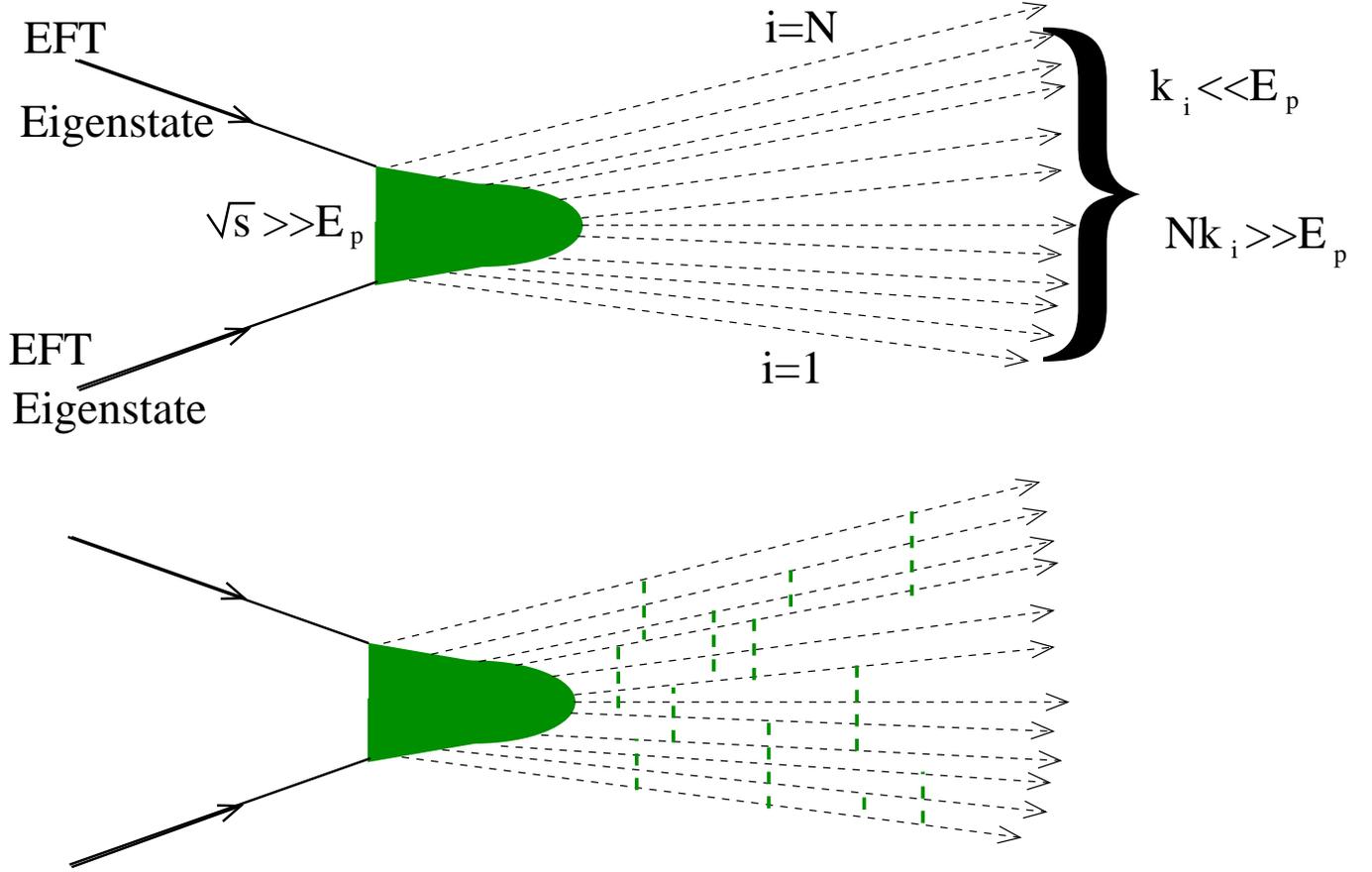}
\caption{\label{bhdiagram} Top: the semiclassical ``scattering diagram'' 
of an event of a hadronic collision of $\sqrt{s} \gg \lqcd$, or of the 
formation of a large evaporating black hole from a pure state  Bottom: the 
Planck suppressed corrections controlling correlations between Hawking 
radiation degrees of fredom.  While these correlations are
suppressed by factors of $(k/l_{planck})^n$ for average occupation 
numbers, they dominate for n-particle cumulants } \end{center}
\end{figure}

Fig. \ref{bhdiagram} top panel summarizes the situation of a black hole 
formed by a coherent quantum state which evaporates into a ``many'' 
particle system. We are bothered by a ``black hole information paradox'' 
due to our conviction that, even though we lack a UV-complete theory of quantum gravity, the {\em effective theory} of quantum gravity relevant to an evaporating black hole is Quantum Field theory on a Schwarzchild background.   The effective theory before and after the black hole exists is, of course, the same quantum field theory on a flat background.
Corrections to these EFTs will necessarily come in factors of $N^{-1}$ and lower.

The essence of the ``black hole information problem'' is that we believe that we can calculate all observables using only the EFT.   The EFT tells us that ``before the black hole formed'', entropy was zero, because we have a pure quantum state, and ``after the black hole formed'' entropy is non-zero because the momentum distribution function seems to be a thermal distribution.   Going from one to another is forbidden in a unitary quantum evolution.

However, the ``non-unitarity of the black hole evolution'' is a necessary artifact of using effective field theory.   The energy of each evaporated particle  $\ave{k}\ll E_P$,  but $E_P \ll M_{BH} \sim \ave{k} N $, where $N$ is the total number of evaporated particles.
Hence, one cannot a priori assume that $N$ particle terms of the density matrix carry an amount of negentropy smaller than 1-particle terms (which set the form of the thermal distribution).   
If the theory of gravity that holds close to the singularity is strongly coupled, as dimensional analysis suggests, the overwhelming fraction of the entropy needs terms at order $(E_p/M_{BH})^{1<n<N}$ to be calculated.
Even if this theory is not strongly coupled, the time required for the black hole to emit most of its entropy $\sim M_{BH}/E_P^2$, so plenty of time should be available for correlations between many many Hawking radiation particles to form.
Such correlations, therefore, will most likely be non-negligible and calculating them {\em cannot be done} using a semiclassical theory with a cutoff, since this theory's approximation {\em is equivalent to assuming these terms be zero} (Fig. \ref{bhdiagram} bottom panel).
For average occupation numbers such terms are suppressed, but not for 
high-order correlations.
Asking ``where the entropy is'' with only the effective theory at our disposal, therefore, is somewhat analogous to asking why the coupling constant of QED runs without having computed a single QED loop diagram, or to take a more everyday example, cooling hot vapor down to ice, and asking why one cannot describe the entropy of the system by an Effective theory of phonons propagating in ice.   
In all these cases, we simply are not using the tools capable of addressing the problem.

``Experimentally'', there might well be non-trivial correlations between ``many'' ($\sim N$) particles emitted from black hole evaporation, which decrease the overall entropy of the $N-$particle final state.   Each correlated state is ``unlikely'' (suppressed by powers of $N^{-1}$), but the number of possible correlations (enhanced by $\sim N!$) beats this unlikelihood.   Their exact nature is only accessible to a theory capable of calculating $(E_p/M_{BH})$ corrections, which of course is lacking at the moment.

In the same way, the resolution of the Firewall paradox pointed out in \cite{braunstein,firewall} might well be that assumption (ii) (as in the abstract of \cite{firewall}) is invalid.
An EFT of n-particle correlations is not expected to conserve unitarity in the IR because entanglement of many final-state particles  is only visible beyond the 1-particle EFT currently at our disposal.   These particles could have been emitted at any time during the black hole evaporation, before or after the absorption of an entangled pair, and still contribute to the correlations balancing the negentropy measured ``long after'' the black hole in question evaporated \cite{entpast1,entpast2}. 
An infalling ``classical'' observer will most likely not see anything special happening at the horizon because, by definition, such a ``classical observer'' does not contain many entangled quantum particles;   As the previous section shows, a correlated quantum pair, of the type examined in works such as \cite{firewall} might well get its entanglement entropy transferred into very complicated correlations of $\sim \order{N}$ particles.
    Conversely, if the observer is some sort of quantum computer with $N \simeq M_{BH}/E_p$ entangled particles, and is governed by dynamics where
such many-particle correlations quickly cascade into 1-particle correlations, they might be able to observe effects similar to what \cite{firewall} call a firewall.
Such carefully prepared, or highly non-linear systems, however, are very far from what we usually consider to be ``an observer''.    Indeed, at present, it is difficult to see, even at the Gedankenlevel, what kind of system would be needed to see the effects required to resolve ``firewall effect''.
Our conclusion is, therefore, very similar to the one suggested in \cite{braunstein}:  the black hole entropy is almost entirely an entropy of entanglement, which will be encoded in complicated inter-particle correlations.   This conclusion (which, as argued by \cite{braunstein}, also preserves the equivalence principle), is unavoidable when the problem is considered from an EFT point of view.

The analysis contained in this work was, by necessity, qualitative and general.
We have made no attempt to describe how ``slow'' quantum dynamics and coarse-graining combine into an effective dynamics which incorporates both unitary and dissipative effects.   
In particular, our arguments cannot as yet be used to isolate the regime where ``fast local thermalization'' can be developed into an effective coarse grained theory of thermalized ``small'' volume elements, i.e. hydrodynamics \cite{kubo,huang}.
We have just guessed what the results might look like in certain limits.
However, as already noted \cite{moore,kovtun,gtfluid,gtfluid2}, the macroscopic gradient expansions usually used to generate effective theories 
(such as hydrodynamics) from fundamental theories have
not one expansion parameter but two:  One, well explored, is the ``Knudsen number'', the ratio of the mean free path to the system size.    The other, the ``microscopic scale'' \cite{dirk} (The mean particle separation $\sim \mathrm{(volume)}\times\mathrm{(density)}^{-1/3}$ in the Boltzmann equation, $\sim \frac{1}{T N_c^{2}}$ in systems with classical gravity duals such as those described in section \ref{secplanar}) is supposed to be parametrically smaller than the mean free path, something which ensures the convergence of the BBGKY expansion.   

The conclusions of this work are similar:
The number of expansion parameters here is also not one but two: The is the coarse-graining scale, $k/\Lambda$ and the state quantum occupancy number $N_{state} \sim (g V \ave{k}^{3})^{-1}$, where $V$ is the total system volume and $g$ the microscopic degeneracy.

If we interpret $\Lambda$ as the inverse mean free path,  we can understand the above two scales in terms of the Density matrix picture.
only when $\Lambda/k \gg N \gg 1$ does the effective theory converge to a Boltzmann equation.    If $N \geq \Lambda/k$, as we saw, the quantum analogue of the BGGKY hierarchy diverges and non-trivial correlations cascade between ``microscopic'' and ``macroscopic'' degrees of freedom.     A ``quantum turbulent'' fluid \cite{gtfluid}, or a weakly coupled plasma of the type studied in \cite{dusling}, might be examples of such systems.
These phenomena are much less studied than the gradient expansion, as we are very far from having analytical tools capable of describing them.   They might, however, be more appropriate for systems with a comparatively small number of degrees of freedom (hadronic collisions, ultracold atoms) than approaches based on either the Boltzmann equation or large $N_c$ expansions.

In conclusion, we have argued that effective field theories are inadequate for describing multi-particle correlations even when the ``fundamental scale'' is well-separated from the probed scale.    We have also argued that, when the theory in the UV is strongly coupled, negentropy is likely to be dominated by such multi-particle correlations.  These arguments must be kept in mind when large non-equilibrium systems are described with effective field theory tools.

We acknowledge the financial support received from the Helmholtz International
Centre for FAIR within the framework of the LOEWE program
(Landesoffensive zur Entwicklung Wissenschaftlich-\"Okonomischer
Exzellenz) launched by the State of Hesse.
GT also acknowledges support from DOE under Grant No. DE-FG02-93ER40764.
This manuscript started as an effort to put down my point of view after some interesting and illuminating discussions with Bill Zajc.   Any misunderstandings in it are however entirely my own doing.
I would also like to thank Mohammed Mia for helpful discussions.

\end{document}